# Solving the Uncapacitated Single Allocation p-Hub Median Problem on GPU

A. Benaini[1], A. Berrajaa[1,2], J. Boukachour[1], M. Oudani[1,3]

1. Normandie UNIV, UNIHAVRE, LMAH, 76600 Le Havre, France.
2. University Mohammed I, LaRI, Oujda, Morocco.
3. University Sidi Mohammed ben abdellah, MSC Lab, Fez, Morocco.

abdelhamid.benaini@univ-lehavre.fr, berrajaa.achraf@gmail.com,
jaouad.boukachour@univ-lehavre.fr, mustapha.oudani@usmba.ac.ma

**Abstract**: A parallel genetic algorithm (GA) implemented on GPU clusters is proposed to solve the Uncapacitated Single Allocation p-Hub Median problem. The GA uses binary and integer encoding and genetic operators adapted to this problem. Our GA is improved by generated initial solution with hubs located at middle nodes. The obtained experimental results are compared with the best known solutions on all benchmarks on instances up to 1000 nodes. Furthermore, we solve our own randomly generated instances up to 6000 nodes. Our approach outperforms most well-known heuristics in terms of solution quality and time execution and it allows hitherto unsolved problems to be solved.

## 1 Introduction

Hubs are sort of facilities that serve to transfer, transhipment and sort in a many-to-many complex distribution networks. They find their applications in airline passengers and fret networks, telecommunications and postal delivery networks. In the air traffic, hubs are the central airports for the long haul by cargo planes for goods and major carriers of passengers. In the telecommunication networks, hubs may be concentrators, routers, multiplexers [26]. In the postal distribution networks, hubs are the major sorting centre and cross docking messaging. The development of this type of network is due to economy of scale achieved by consolidating the traffic through the hub-hub arcs [1].

A rich scientific literature about hub location problems has been developed since 1980 and articles number has increased recently. Different variants of hub location problems have been defined and classified according to allocation way: the single allocation where each spoke (non-hub node) is assigned to exactly one hub and the multiple allocation that enables the spokes to be allocated to several hubs. The p-hub median problem when the number p of hubs to be located is given otherwise the problem is hub location. According to hubs capacities, the problem is said to be Uncapacitated (resp. capacitated) if hubs have infinite (resp. finite) capacities. There are several other kinds of hub problems like the p-hub centre problem where the objective is to minimize the maximum travel time between two demand centres [8], the hub arc problem which aims to overtake the shortcoming of the p-hub median problem by introducing the bridges arcs between hubs without discount factor [9], the dynamic hub location problem where either cost, demands or resources may vary in the planning horizon [12]. Other constraints can be taken into account, such as, hubs congestion, non-linear costs, stochastic elements, or vehicles routing constraints [14]. Reviews, synthesis and classification on models and methods used in literature on different variants of the hub location problem can be found in [4], [10], [19], [23], [30].

This paper deals with the Uncapacitated Single Allocation p-Hub Median Problem (USApHMP) for which we propose a parallel GA approach on GPU. To our knowledge, this is the first parallel GPU implementation for solving this problem. The remainder of this paper is organized as follows, related works are provided in Section 2. In Section 3, we present the mathematical formulation of the problem. The parallel GA approach is described in Section 4 followed by the GPU implementation in Section 5. Computational results are reported in the Section 6 and finally Conclusion and perspectives are given in Section 7.

This work was partially supported by the Region Normandy under CLASSE project.

## 2  Related Works

O'kelly et al. [31] presented the first mathematical formulation for the USApHMP as a quadratic integer program. They developed two heuristics and reported numerical results for CAB data (Civilian Aeronautics Board) with 25, 20, 15 and 10 nodes. Campbell et al. [8],[9] presented different formulations for the p-hub median problem, the uncapacitated hub location problem, the p-hub center problem and the hub covering problem. Possible extensions with flow thresholds are also studied. They introduced the p-hub median problem and proposed two heuristics to handle instances with 10–40 nodes and up to 8 hubs. Skorin-Kapov et al. [35], developed different mixed 0–1 linear formulations for the multiple and the single p-hub median problems and reported results on CAB data set. Sohn and Park [36], studied the special case of single allocation two-hubs location problem. In this particular case, the quadratic program is transformed to a linear program and to a minimum cut problem. Abdinnour-Helm [1] proposed a hybrid GA and tabu search heuristic and reported the results on the CAB data set. Ernst and Krishnamoorthy [16], presented a solving approach to the multiple allocation p-hub median problem and described how the approach can be adapted to the single allocation case. Results are reported on AP data for multiple allocation case up to 200 nodes. Bryan [7] studied four hub-and-spoke networks. The first is concerned by capacitated network, the second focus in minimum threshold model, the third determines the numbers of open hubs and the last introduce flow-dependent cost function.

Horner and O'Kelly [21] proposed a model implemented in a GIS environment to prove that hub networks may emerge naturally on traffic networks to take advantages of economies of scale. Labbé et al. [26] studied the polyhedral properties of the single assignment hub location problem and proposed a Branch-and-Cut algorithm for solving this variant of hub location. Chen [11] proposed a hybrid heuristic to solve the USAHLP based on a combination of an upper bound method search, simulated annealing and tabu list heuristic. Tests were performed on CAB data and AP data up to 200 nodes. Silva and Cunha [34] proposed three variants of tabu search heuristics and a two-stage integrated tabu search to solve the problem. The authors used the multi-start principle to generate different initial solutions which are improved by tabu search. They solved larger instances with 300 and 400 nodes. Ilic et al. [22] proposed a general variable neighborhood search for the USApMLP. They reported the results on AP and PlanetLab instances and Urand instances up to 1000 nodes. de Camargo and Miranda [14], introduced the single allocation hub location problem under congestion. A generalized Benders decomposition algorithm is proposed to solve AP instances.

Maric et al. [27] proposed a memetic algorithm based on two local search heuristics. They tested their algorithm on the well-known benchmarks and created larger scale instances with 52–900 nodes. They gave the optimal solutions of AP data up to 200 nodes. Bailey et al. [5] proposed a Discrete Particle Swarm Optimization (DPSO) to solve the USAHLP. They obtained the optimal solutions on all CAB data set and on AP data up to 200 nodes. Damgacioglu et al. [13] introduced a planar version of the uncapacitated hub single allocation hub location problem. This version has the particularity that a hub can be located anywhere in the plan. They reported the results on benchmarks AP data instances. Ting and Wang [38] proposed a threshold accepting TA algorithm to solve the USAHLP and reported results on the AP and CAB benchmarks. Meier and Clausen [28] made use of the data set structures to propose new linearization of the quadratic formulation of the problem. Indeed, the Euclidean distance in instances enabled to get linearization of three classical and two new formulations of the single allocation problem. They obtained optimal solutions on the AP data up to 200 nodes. Rostami et al. [33] introduced a new version of the USApHMP where the discount factor between hubs representing scale economy in hub-hub arcs is replaced by a decision variable. They proposed a Branch-and-bound algorithm and Lagrangian relaxation to compute lower bounds. Recently, Abyazi-Sani and Ghanbari [2] proposed a Tabu Search heuristic for solving the USAHLP and reported the results both on CAB data and AP data set up to 400 nodes. Kratica [25] proposed a GA for solving the uncapacited multiple allocation hub problem. Binary encoding and adapted genetic operation to this problem are used (only allocation hubs are given as the solution). He shows, under experimental results on ORLIB instances with up to 200 nodes that GA approach quickly reaches all optimal solution that are known. Topcuoglu et al. [39] present a GA approach to solve the uncapacited hub location problem. We use their encoding and GA operators in our parallel GA. However, we generate initial solutions differently from the middle nodes (rather than randomly initial solution as in [39]) with aiming to reach more quickly the best solutions.

Parallel GA implementations have been the subject of many works. There is extensive emerging research in this field and several studies suggest different strategies to implement GAs on different parallel machines [3], [18], [20], [24], [32], [36], [37]. There are three major types of Parallel GAs: (1) the master-slave model, (2) island model and (3) fine-grained model. In the master-slave model, the master node holds the population and performs most of the GA operations. The fitness evaluation, the crossover, the correction and mutation operations on groups of individuals are made by each slave. In a coarse-grained model, the population is divided into several nodes. Each node then has a subpopulation on which it executes GA operations. In fine-grained models, each node only has a single individual, and each node can only communicate with several neighboring nodes. In this case, the population is the collection of all the individuals in each node. There are conflicting reports over whether multiple independent runs of GAs with small populations can reach solutions of higher quality or can find acceptable solutions faster than a single run with a large population.

In this work, we propose GPU implementation of GA for solving the USApHMP. Several GPU implementations of parallel GA are proposed in the literature. Among them, [6], [32] presented the mapping of the parallel island-based GA on GPU. Our approach is similar to these implementations, nevertheless, the migration step is replaced by a selection of the best solutions in each iteration, and the generation of the initial solution is quite different (from the middle nodes).

## 3   Problem formulation

The USApHMP can be stated as follows: given N nodes 1…N, we try to locate p hubs and to find an optimal allocation of spokes to hubs (one hub for each spoke) that minimizes the sum of the total flow cost.

Let $Z_{ik}$ be the binary decision variable equal to 1 if the node i is assigned to the hub k, 0 otherwise, $Y_{kl}^i$ the flow between the hubs k and l originated from the node i, $C_{ik}$ the unit cost for the flow in the arc (i,k), $O_i$ and $D_i$ are the originated and destined flow to the node i respectively.

The USApHMP is formulated as a MIP (Mixed Integer Program) by Ernst and Krishnamoorthy [15] as follows:

$$minimize \sum_i \sum_k C_{ik} Z_{ik} (\chi O_i + \delta D_i) + \sum_i \sum_k \sum_l \alpha C_{kl} Y_{kl}^i \qquad (1)$$

Subject to:

$$\sum_k Z_{ik} = 1, \forall\, i \in N \qquad (2)$$

$$Z_{ik} \leq Z_{kk}, \forall\, i,k \in N \qquad (3)$$

$$\sum_l Y_{kl}^i - \sum_l Y_{lk}^i = O_i Z_{ik} - \sum_j W_{ij} Z_{jk}, \forall\, i,k \in N \qquad (4)$$

$$\sum_k Z_{kk} = p \qquad (5)$$

$$Z_{ik} \in \{0,1\}, \quad 1 \leq i,k \leq N$$
$$Y_{kl}^i \geq 0, \quad 1 \leq i,l,k \leq N$$

The objective function (1) minimizes the total cost of flow transportation between all origin-destination nodes. Constraint (2) imposes to each spoke to be assigned to exactly one hub (additionally each hub is allocated to itself). Constraint (3) requires that spokes will be assigned to hubs if the last one were open. Constraint (4) is the flow conservation constraint and constraint (5) imposes to locate exactly p hubs.

The USApHMP is known to be NP-hard with exception of special cases that are solved in polynomial time. When the set of hubs is fixed then the problem can be solved in $O(n^3)$ time using the shortest-path algorithm [17].

## 4   Genetic algorithm description

Genetic algorithms are well-known search approaches that are applied in the wide field of optimization. So, we propose a parallel GA to solve the USApHMP on GPU. Our implementation quickly reaches the optimal or best solutions for all benchmarks. In the next subsection we detail the encoding chromosomes and how generate the initial solution of this problem.

### 4.1 Encoding and initial solution

Each solution of the problem is represented by two N-arrays H and S (this encoding was used in [39]) where:
- H represents hub locations i.e H[i]= 1 if node i is a hub, H[i]=0 otherwise.
- S represents the allocation of spokes (non-hub nodes) to hubs i.e S[i]= k where k is the assigned hub for the node i. Additionally, each hub is allocated to itself.

In [39], the initial solution is generated pseudo randomly. Here, we proceed differently in order to quickly achieve best hubs locations. So to build an initial solution with p hubs, we first compute the p middle nodes i.e. the p hubs i with smallest distances $d_i$ to the other nodes with $d_i = \sum_j C_{ij}$. Thus, the p initial hubs are chosen among the p middle nodes. Then each node is allocated to its nearest hub.

**Numerical example:**

The Fig .1 shows an example of a solution with 7 nodes, 2 hubs (nodes 2 and 5). The nodes 1, 2 and 6 are allocated to hub 2 and the other nodes are allocated to node 5. The encoding of this solution is given in Fig. 1:

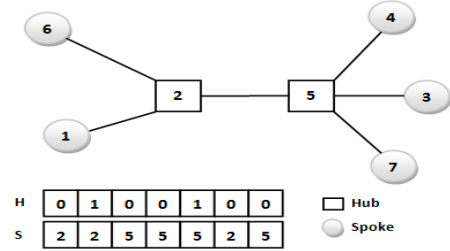

The initial population is generated by duplication of the initial solution by randomly permuting one hub with one spoke.

**Fig. 1**: simple network encoding

### 4.2 Genetic operators

Random single point-crossover operator is used and infeasible offspring are corrected by a specific operator to ensure validity of solutions, in terms of number of hubs by assigning the corresponding spokes to their neighbor hubs. The permutation of two hubs is used as a mutation operator. These operators are noted crossover(), correction() and mutation() respectively.

### 4.3 Solution evaluation

The following Eval() definitions (fitness) are used in the standard benchmarks to evaluate the solutions quality. The fitness version for CAB data is given by:

$$(\sum_i \sum_k C_{ik} Z_{ik}(\chi O_i + \delta D_i) + \sum_i \sum_k \sum_l \alpha C_{kl} Y_{kl}^i) * 1/ \sum_i \sum_j W_{ij}$$

The fitness version for all data instances except PlanetLab and CAB data is given by:

$$10^{-3}(\sum_i \sum_k C_{ik} Z_{ik}(\chi O_i + \delta D_i) + \sum_i \sum_k \sum_l \alpha C_{kl} Y_{kl}^i)$$

Note that the reason to multiply by $10^{-3}$ is to obtain the unit cost for flow transportation. We discover this when we tried to reproducing optimal solutions and we confirm it by contacting M.R Silva [34].

## 5 GPU implementation

The Graphics Processing Units are now available in most of personal computers. They are used to accelerate the execution of variety of problems. The smallest unit in GPU that can be executed is called thread. Threads (all executing the same code and can be synchronized) are grouped into blocks of equally sized and blocks are grouped in grid (blocks are independent and cannot be synchronized).

The memory hierarchy of the GPU consists of three levels: 1) the global memory that is accessible by all threads. 2) the shared memory accessible by all threads of a block and 3) the local memory (register) accessible by a thread. Shared memory has a low latency (2 cycles) and is of limited size. Global memory has a high latency (400 cycles) and is of large size (4 GB for the Quadro). An entire block is assigned to a single SM (Stream Multiprocessor). Each SM is composed of 32 streaming processors that share a limited size shared memory. Several blocks can run on the same SM. Each block is divided into Warps (32 threads by Warp) that are executed in parallel. The programmer must control the block sizes, the number of Warps and the different memories access.

A typical CUDA program is a C program where the functions are distinguished based on whether they are meant for execution on the CPU or on the GPU. The functions executed on the GPU are called kernels and are executed by several threads. We implemented the GA on GPU (Nvidia Quadro with 4 GB and 384 cores running under CUDA 7.5 environment) and we compare it to sequential implementations of best known results existing articles in the literature in terms of time computations and on solutions quality. We showed the effectiveness of our implementation on several instances of the USApHMP.

Fig. 2 gives the schema of the parallel GA implementation on GPU. The following parameters are used: The number of node N, the population size n, the number of generations R, the number of iterations in the inner-loop N1, the number of iterations in the outer-loop N2.

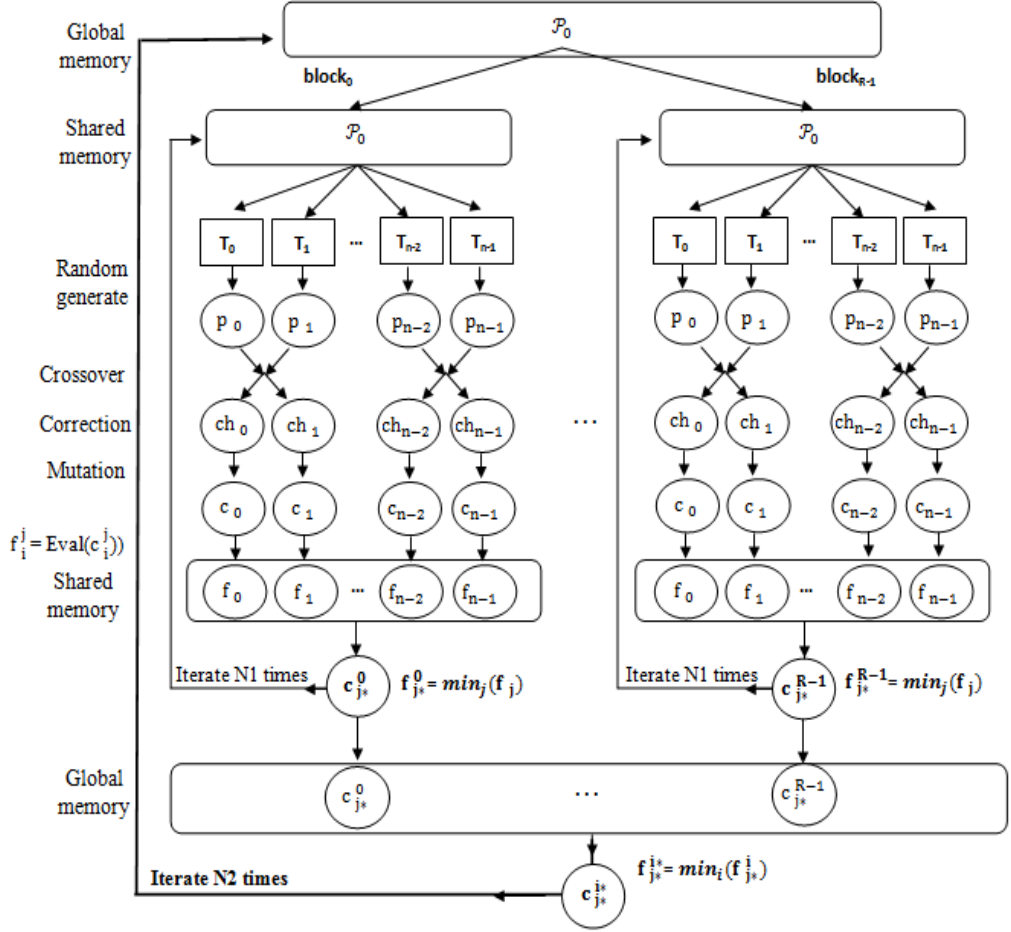

Fig. 2 The schema of the parallel GPU implementation of the GA.

We partition the GPU on R blocks each one is a gird n x 1 of threads. The master thread of each block is the thread 0 and the global master thread is the thread 0 of the block 0. The block i, $0 \leq i < R$ stores in its shared memory the data required to execute one GA starting from an ancestor individual $\mathcal{P}_0$ (initial feasible solution) generated as indicated in section 3 by the CPU and copied in the global memory of the GPU.

More precisely, let $T_0^i, \ldots, T_{n-1}^{R-1}$ the threads of the block i. Starting from $\mathcal{P}_0^i$, each thread $T_j^i$ generate a new solution (individual) $p_j^i$ by applying a random permutation to $\mathcal{P}_0^i$ (initially, $\mathcal{P}_0^i = \mathcal{P}_0$ and is updated after each iteration of the inner-loop). $\mathcal{P}_0^i = p_0^i, \ldots, p_{n-1}^i$ is the initial population of the GA executed by the block i. Note that the population size n is the same for all the blocks.

Now, we explain how the block i executes the GA. Each thread $T_{2j}^i$ of the block i generate two children namely $ch_1$ and $ch_2$ by crossowing the parents $p_{2j}^i$, $p_{2j+1}^i$ then $T_{2j}^i$ applies the mutation to $ch_1$ to get a new individual say $c_{2j}^i$ and $T_{2j+1}^i$ applies the mutation to $ch_2$ to get a new individual say $c_{2j+1}^i$. Next, each thread

$T_j^i$ executes the correction operator to ensure the validity of the solution, verifies that all nodes are assigned to the nearest hubs and finally it computes $f_j^i = Eval(c_j^i)$.

Note that all $c_j^i$ and $f_j^i$ are stored in the shared memory of the block i. So, the master thread of block i selects the individual $c_{j*}^i$ with $min_j\{f_j^i\}$ and updates the ancestor $s_0^i$ as $c_{j*}^i$ for the next iteration. This inner-loop of GA terminates after N1 iterations (the same for all the blocks).

The $c_{j*}^i$, $0 \leq i < R$, are copied in the global memory and the individual $c_{j*}^{i*}$ with the $min_i\{f_{j*}^i\}$ is selected as the final solution or as the new value of the ancestor $s_0$ for the next iteration of the outer-loop. The process is repeated N2 times.

The pseudo CUDA code executed by the CPU is the following:
1. Generate($\mathcal{P}_0$); //the ancestor individual
2. Copy $\mathcal{P}_0$ in the global memory of the GPU.
3. Define the blocks and the grid :
   dim3 dimBlock(n,1);
   dim3 dimGrid(R,1);
4. Launch the kernel GA($\mathcal{P}_0$) : GA<<<dimGrid,dimBlock>>>($\mathcal{P}_0$);
5. Read the solution from the global memory.

# 6   Computational results

### 6.1 Benchmarks used

We used four types of data: CAB, AP, PlanetLab and Urand :

- CAB (Civilian Aeronautic Board) data set is set of instances introduced in [31] based on airline passenger flow between 25 US cities. It contains distances (which satisfy triangle inequality) and symmetric flow matrix between the cities. The size instances are of 10, 15, 20 and 25 nodes. The distribution and collection factors δ and χ are equal to 1.

- AP (Australian Post) data set are real-world data set representing mail flows in Australia. The distribution and collection factor δ and χ equal 3 and 2 respectively while the discount factor α takes 0.75 for all instances. The mail flows are not symmetric and there are possible flows between each node and itself.

- Urand data set are random instances up to 400 nodes generated by Meyer et al. [29]. The instances with 1000 nodes were generated by Ilic et al. [22]. Nodes coordinates were randomly generated from 0 to 100000 and the flow matrix was randomly generated.

- The PlanetLab instances are node-to-node delay information for performing Internet measurements [22]. In these networks, χ = α = δ = 1 and the distance matrix doesn't respect triangle inequality.

### 6.2 Best known solutions vs. our results

We report the results for the three data set introduced above. We compare our results with those of Ilic et al. [22] in terms of computing time. Note that in our GPU implementation, the number of blocks is the same for all problems. So time compute of all problems is the same. We use shared memory to reduce the time computation. However, the time transfer between the CPU and GPU varies according to the number of nodes. Throughout the rest the given times are the time of the complete program (calculation of the initial solution, data transfers between the CPU and GPU, calculation of the solution).

For the CAB data set, we obtained the optimal solutions in all instances (up to 25 nodes) in a short computing time. Since solving these instances is not anymore a challenge (all instances are solved to optimality by previous work), we report only our computing times for solving these data instances. We studied the scale economy generated in hub-hub arcs and its relationship with initial and final costs which represent the collection and distribution cost. Typically, a hub-and-spoke transportation chain is composed by tree segments: the first and the third called pre and post haul respectively are the initial and final arcs while the second is the long haul segment (hub-hub arcs). In CAB data, we can express the cost from an origin i to a destination j through the two hubs k and l as: $\boldsymbol{C_{ij}^{kl} = \chi C_{ik} + \alpha C_{kl} + \delta C_{jl}}$ where α ≤ 1 represents the scale economy generated by consolidating flows between hubs while χ and δ represent the distribution and the collection costs and are often greater than 1. A question is how distribution and collection cost influence the scale economy thresholds? In fact, as illustrated in Fig. 3, the average inter-hub distance changes as we vary the distribution and collection factors. We can see clearly that the long-haul relevance threshold is lower when the distribution and collection costs are lower.

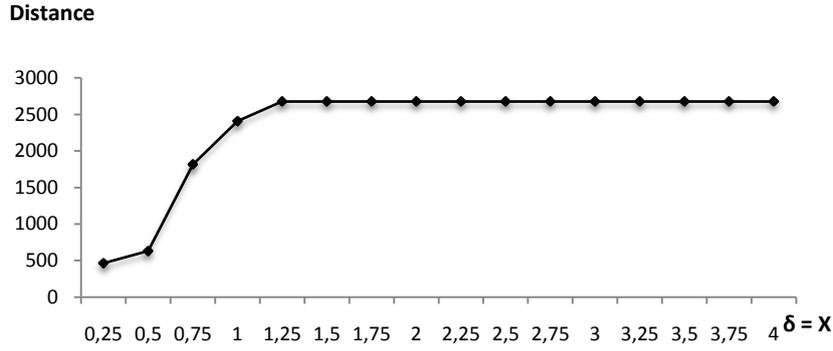

Fig. 3 The average inter-hubs distance

The following notations are used in Tables 1-5:
- N: nodes number in the instance.
- p: hubs number.
- Best Sol: the Best solution if it is known otherwise "-"is written.
- GPU Sol: the best solution obtained by GPU, with mark "opt" when solution in GPU is the optimum for the current instance.
- $T_{OPT}$ : the best time (in seconds) for the best solution.
- $T_{GPU}$ : the time (in seconds) for our GA parallel.

These tables give a comparison of our results to the best known results for USApHMP in the benchmarks previously introduced. As shown in Table 1 we obtained optimal solutions for all the AP data instances in time ≤ 7.42 s. Note that, the results using AP data instances for the p-hub median variant with 300 and 400 nodes are not reported before in the literature and we think that finding exact solutions using standard solvers (CPLEX, Gurobi...) is a serious challenge. So, we can think that our results are since now the best solutions for 300 and 400 nodes instances. We report our results for PlanetLab instances in Table 2. It is clear that our approach outperforms those of literature [22] either in cost and computing time. The state of the art solutions given in [22] reports the results for 12 instances. Each instance is characterized by nodes number n and by p hubs to be located with $p \approx \sqrt{n}$.

**Table 1: Results on AP data**

| N | p | Best Sol | GPU Sol | $T_{GPU}$ | N | p | Best Sol | GPU Sol | $T_{GPU}$ |
|---|---|---|---|---|---|---|---|---|---|
| 10 | 2 | 167493.06 | opt | 0.007 | 100 | 5 | 136929.444 | opt | 1.310 |
|  | 3 | 136008.13 | opt | 0.012 |  | 10 | 106469.566 | opt | 1.310 |
|  | 4 | 112396.07 | opt | 0.014 |  | 15 | 90533.523 | opt | 1.49 |
|  | 5 | 91105.37 | opt | 0.019 |  | 20 | 80270.962 | opt | 1.63 |
| 20 | 2 | 172816.69 | opt | 0.020 | 200 | 5 | 140062.647 | opt | 3.602 |
|  | 3 | 151533.08 | opt | 0.031 |  | 10 | 110147.657 | opt | 3.722 |
|  | 4 | 135624.88 | opt | 0.039 |  | 15 | 94459.201 | opt | 3.783 |
|  | 5 | 123130.09 | opt | 0.043 |  | 20 | 84955.328 | opt | 3.841 |
| 25 | 2 | 175541.98 | opt | 0.033 | 300 | 5 | - | **174914.73** | 5.631 |
|  | 3 | 155256.32 | opt | 0.045 |  | 10 | - | **134773.55** | 5.711 |
|  | 4 | 139197.17 | opt | 0.050 |  | 15 | - | **114969.85** | 5.896 |
|  | 5 | 123574.29 | opt | 0.061 |  | 20 | - | **103746.44** | 5.876 |
| 40 | 2 | 177471.67 | opt | 0.063 | 400 | 5 | - | **176357.92** | 6.741 |
|  | 3 | 158830.54 | opt | 0.110 |  | 10 | - | **136378.19** | 6.846 |
|  | 4 | 143968.88 | opt | 0.167 |  | 15 | - | **117347.10** | 7.102 |
|  | 5 | 134264.97 | opt | 0.213 |  | 20 | - | **104668.27** | 7.423 |
| 50 | 2 | 178484.29 | opt | 0.092 |  |  |  |  |  |
|  | 3 | 158569.93 | opt | 0.163 |  |  |  |  |  |
|  | 4 | 143378.05 | opt | 0.250 |  |  |  |  |  |
|  | 5 | 132366.953 | opt | 0.271 |  |  |  |  |  |

**Table 2: Results on PlanetLab**

| Instance | N | p | Best Sol | GPU Sol | $T_{OPT}$ | $T_{GPU}$ |
|---|---|---|---|---|---|---|
| 01-2005 | 127 | 12 | 2927946 | **2904434** | 148.954 | **0.47** |
| 02-2005 | 321 | 19 | 18579238 | **18329984** | 462.790 | **6.95** |
| 03-2005 | 324 | 18 | 20569390 | **20284132** | 543.844 | **7.54** |
| 04-2005 | 70 | 9 | 739954 | **730810** | 0.682 | **0.28** |
| 05-2005 | 374 | 20 | 25696352 | **25583240** | 622.612 | **8.32** |
| 06-2005 | 365 | 20 | 22214156 | **22191592** | 581.776 | **7.94** |
| 07-2005 | 380 | 20 | 30984986 | **30782956** | 546.688 | **8.47** |
| 08-2005 | 402 | 21 | 30878576 | **30636170** | 637.686 | **8.74** |
| 09-2005 | 419 | 21 | 32959078 | **32649752** | 684.900 | **9.34** |
| 10-2005 | 414 | 21 | 32836162 | **32687796** | 731.930 | **9.12** |
| 11-2005 | 407 | 21 | 27787880 | **27644374** | 588.344 | **9.22** |
| 12-2005 | 414 | 21 | 28462348 | **28213748** | 680.382 | **9.18** |

**Table 3: Results on Urand instances**

| N | p | Best Sol | GPU Sol | $T_{GPU}$ | N | p | Best Sol | GPU Sol | $T_{GPU}$ |
|---|---|---|---|---|---|---|---|---|---|
| 100 | 2 | 36930.31 | opt | 0.0375 | 300 | 2 | 328702.42 | opt | 0.2215 |
| | 3 | 34532.88 | opt | 0.0265 | | 3 | 308765.08 | opt | 0.9175 |
| | 4 | 32608.28 | opt | 0.0245 | | 4 | 293636.81 | opt | 1.6100 |
| | 5 | 31107.70 | opt | 0.1135 | | 5 | 282116.88 | opt | 0.5060 |
| | 10 | 27058.40 | opt | 0.4695 | | 10 | 251393.30 | opt | 12.8275 |
| | 15 | 25408.56 | opt | 2.7925 | | 15 | 236781.77 | opt | 27.6640 |
| | 20 | 24377.65 | opt | 7.3640 | | 20 | 228005.19 | opt | 153.2925 |
| 200 | 2 | 148235.45 | opt | 0.0175 | 400 | 2 | 579982.35 | opt | 0.1735 |
| | 3 | 139223.25 | opt | 0.0575 | | 3 | 543717.32 | opt | 1.7115 |
| | 4 | 132676.89 | opt | 0.3920 | | 4 | 519217.48 | opt | 2.2275 |
| | 5 | 127220.02 | opt | 0.6895 | | 5 | 501421.52 | opt | 1.4730 |
| | 10 | 112539.21 | opt | 4.5300 | | 10 | 446361.10 | opt | 16.8700 |
| | 15 | 105690.52 | opt | 37.3460 | | 15 | 422284.78 | opt | 111.4295 |
| | 20 | 102022.32 | opt | 68.6685 | | 20 | 407110.51 | opt | 228.8615 |

**Table 4: Results on Urand large instances**

| N | p | Best Sol | GPU Sol | $T_{OPT}$ | $T_{GPU}$ |
|---|---|---|---|---|---|
| 1000 | 2 | 198071412.53 | **8184986.50** | 1.7245 | 9.321 |
| | 3 | 169450816.35 | **7024184.00** | 8.1550 | 9.785 |
| | 4 | 150733606.87 | **6184749.01** | 2.2240 | 10.431 |
| | 5 | 142450250.26 | **5860994.06** | 58.6070 | **10.89** |
| | 10 | 114220373.07 | **4752317.00** | 187.8385 | **13.7** |
| | 15 | - | **4228256.88** | - | 15.23 |
| | 20 | 198071412.53 | **3928617.48** | 403.4280 | 17.923 |

**Table 5: Results on larger Urand instances (generated by us)**

| N | p | GPU Sol | $T_{GPU}$ | N | p | GPU Sol | $T_{GPU}$ |
|---|---|---|---|---|---|---|---|
| **1500** | 20 | **454787506** | 196 | **4000** | 20 | **3234999192** | 3076 |
| | 30 | **407155164** | 286 | | 30 | **2983891783** | 3276 |
| | 40 | **380114045** | 423 | | 40 | **2769550514** | 3365 |
| | 50 | **363586538** | 574 | | 50 | **2644606684** | 3648 |
| **2000** | 20 | **805749722** | 477 | **5000** | 20 | **5085803132** | 4662 |
| | 30 | **733375448** | 580 | | 30 | **4656787498** | 4720 |
| | 40 | **686515363** | 714 | | 40 | **4353561395** | 4996 |
| | 50 | **655938000** | 965 | | 50 | **4143849388** | 5112 |
| **3000** | 20 | **1804950952** | 1157 | **6000** | 20 | **7398401957** | 5614 |
| | 30 | **1642145354** | 1544 | | 30 | **6675723961** | 5748 |
| | 40 | **1538548764** | 1869 | | 40 | **6293053841** | 5964 |
| | 50 | **1468780124** | 2086 | | 50 | **5999780197** | 6212 |

The Table 3 reports computational results for the Urand instances. We can see that our parallel GA obtained the best solutions for instances up to 400 nodes and outperforms those of Ilic et al., [22] for instances with 1000 nodes as illustrated in Table 4. Concerning the computing times, our approach is faster and gets the solutions in a time lapse less than 18s for all instances while the best-known time reaches 7 minutes. A remarkable thing is that the time execution gap of our algorithm with Ilic et al., [22] algorithm is important for large values of p. We report in Table 5 results for larger instances generated by us using the same generation procedure as for the Urand instances as stated in [29]. These new challenging instances consist of large networks up to 6000 nodes that have not been solved before.

# 7    Conclusion and perspectives

We developed a parallel GA for the Uncapacitated Single Allocation p-Hub Median problem and we implement it on GPU. We showed the effectiveness of our implementation on the well-known benchmarks for this problem. Indeed, our approach improved the best known solutions in cost and computing times for well-known benchmarks instances with up to 1000 nodes. Also it allowed solving large instances problem unsolved before. Further, we work on the design and implementation of an exact parallel tree-based algorithm to solve the studied hub problems as these algorithm structures seems to be suitable for the GPU architectures. Another issue is to tackle other versions of the hub problem especially capacitated case, multiple allocation variants and other more specific problems (with congestion, with vehicles routing constraints, etc.). Other metaheuristics in particular those based on one solution may be studied from the parallelism viewpoint.

**Acknowledgments**

We thank Dr, M. O'Kelly, M.R. Silva, C.B. Cunha, Ilic A and R. Abyazi-Sani for instances data set provided.